\documentclass[aps,prl,twocolumn,superscriptaddress,nobibnotes,amsmath,amssymb]{revtex4-1}

\usepackage{graphicx}
\usepackage{color}
\usepackage[explicit]{titlesec}
\usepackage[normalem]{ulem}

\begin{document} 
\title{Topological protection brought to light by the time-reversal symmetry breaking}
\author{S.U.~Piatrusha}
\affiliation{Institute of Solid State Physics, Russian Academy of Sciences, 142432 Chernogolovka, Russian Federation}
\affiliation{Moscow Institute of Physics and Technology, Dolgoprudny, 141700 Russian Federation}
\author{E.S.~Tikhonov}
\email[e-mail:]{tikhonov@issp.ac.ru}
\affiliation{Institute of Solid State Physics, Russian Academy of Sciences, 142432 Chernogolovka, Russian Federation}
\affiliation{Moscow Institute of Physics and Technology, Dolgoprudny, 141700 Russian Federation}
\author{Z.D.~Kvon} 
\affiliation{Institute of Semiconductor Physics, Novosibirsk 630090, Russian Federation}
\affiliation{Novosibirsk State University, Novosibirsk 630090, Russian Federation}
\author{N.N.~Mikhailov}
\affiliation{Institute of Semiconductor Physics, Novosibirsk 630090, Russian Federation}
\affiliation{Novosibirsk State University, Novosibirsk 630090, Russian Federation}
\author{S.A.~Dvoretsky}
\affiliation{Institute of Semiconductor Physics, Novosibirsk 630090, Russian Federation}
\author{V.S.~Khrapai}
\affiliation{Institute of Solid State Physics, Russian Academy of Sciences, 142432 Chernogolovka, Russian Federation}
\affiliation{Moscow Institute of Physics and Technology, Dolgoprudny, 141700 Russian Federation}

\date{\today}	

\begin{abstract}
Recent topological band theory distinguishes electronic band insulators with respect to various symmetries and topological invariants, most commonly, the time reversal symmetry and the $\rm Z_2$~invariant. The interface of two topologically distinct insulators hosts a unique class of electronic states -- the helical states, which shortcut the gapped bulk and exhibit spin-momentum locking. The magic and so far elusive property of the helical electrons, known as topological protection, prevents them from coherent backscattering as long as the underlying symmetry is preserved. Here we present an experiment which brings to light the strength of topological protection in one-dimensional helical edge states of a~$\rm Z_2$ quantum spin-Hall insulator in HgTe. At low temperatures, we observe the dramatic impact of a tiny magnetic field, which results in an exponential increase of the resistance accompanied by giant mesoscopic fluctuations and a gap opening. This textbook Anderson localization scenario emerges only upon the time-reversal symmetry (TRS) breaking, bringing the first direct evidence of the topological protection strength in helical edge states.
\end{abstract}

\maketitle

Bloch's quantum mechanical band theory states that electronic waves propagate freely in a periodic lattice of a crystalline solid~\cite{Bloch1929}. A defect of any kind breaks the translational symmetry of the lattice and mediates scattering, similar to the diffraction of plane waves in a defected continuous medium. With increasing the number of defects, the free propagation of a wave turns into a diffusion. Given the phase coherence is preserved, the constructive interference between the time-reversed random diffusion paths gives rise to a coherent backscattering of a wave. This genuine quantum effect is observed as a narrow resonance in the intensity of light backscattered off a milky solution~\cite{Kuga1984,PhysRevLett.55.2692,PhysRevLett.55.2696} and as a weak localization correction to the conductance of a diffusive metal~\cite{nazarov_blanter_2009}. Eventually, in a sufficiently disordered system, the coherent propagation gets suppressed, be it light~\cite{Wiersma1997,Schwartz2007}, electron~\cite{RevModPhys.57.287}, or even sound~\cite{Hu2008} or matter waves~\cite{Roati2008,Billy2008} -- the phenomenon known as Anderson localization~\cite{PhysRev.109.1492,PhysRevLett.42.673}. For the helical states, however, the situation inverts thanks to a destructive interference of the time-reversed paths. Thereby the free propagation is maintained in the presence of a symmetry-conserving disorder, which is known as topological protection~\cite{RevModPhys.82.3045,RevModPhys.83.1057}.

Helical edge states represent a unique example of a 1D electronic system, that can only be realized at the interface of a two-dimensional (2D) $\rm Z_2$ topological and trivial insulators~\cite{RevModPhys.83.1057}, in this work -- the quantum spin-Hall (QSHI) insulator in HgTe/CdHgTe quantum wells (QWs) with the inverted band structure and vacuum~\cite{Bernevig1757}. Spin-momentum locking is manifested in two counter-propagating opposite-spin species at each edge of a planar device, which provide the only transport channel when the Fermi level is tuned within the 2D bulk energy gap~\cite{Konig2007}. Numerous direct consequences of this physical picture are corroborated experimentally, including the observation of quantized conductance~$G\approx G_0\equiv e^2/h$ of the shortest edge channels~\cite{Konig2007,PhysRevLett.107.136603,Olshanetsky2015}, non-local transport in zero magnetic field~\cite{Roth2009,PhysRevB.87.235311,Olshanetsky2015,Li2017,Fei2017,Piatrusha2017}, positive magnetoresistance~\cite{Konig2007,Gusev2013}, the spin-charge sensitivity~\cite{Brune2012} and the unconventional behavior in lateral p-n junctions~\cite{Minkov2015,Piatrusha2017}. 

In spite of the impressive progress, the mean-free path of the helical electrons is typically disappointingly small~\cite{Konig2007,Roth2009,Gusev2011,PhysRevB.87.235311,Nowack2013,Olshanetsky2015,PhysRevLett.114.096802,Fei2017,Li2017,Wu2018,Bendias2018}, even compared to the conventional high-purity 1D conductors~\cite{dePicciotto2001}. The puzzles of the trivial ohmic behavior and weak or even absent temperature dependence~\cite{Konig2007,Gusev2011,Nowack2013,Gusev2014,Olshanetsky2015,PhysRevLett.114.096802,Tikhonov2015,Wu2018,Bendias2018} along with the nearly universal partition noise~\cite{Tikhonov2015,Aseev2016,Piatrusha2018} further indicate that the edge transport beyond the mean-free path in zero magnetic field is classical, rather than quantum coherent, by nature. As a matter of fact, the advantages offered by the concept of topological protection, which are of paramount importance for numerous applications~\cite{Qian1344,PhysRevLett.100.096407}, were so far hidden by an extremely efficient phase breaking mechanism of a debated origin~\cite{Xu2006,PhysRevLett.102.256803,Vayrynen2013,Kainaris2014,Wang2017,Vayrynen2018,PhysRevLett.122.016601}. Here, we approach this problem from a different perspective, using coherent backscattering as a marker for a breakdown of the topological protection. We observe that the TRS breaking by magnetic field restores the coherent backscattering, thereby drastically decreasing the mean-free path in a finite magnetic field, and drives the Anderson localization of the helical edge channels. This behavior unveils 
the actual strength of the topological protection in the TRS case.

\begin{figure*}
	\begin{center}
		\includegraphics[scale=1]{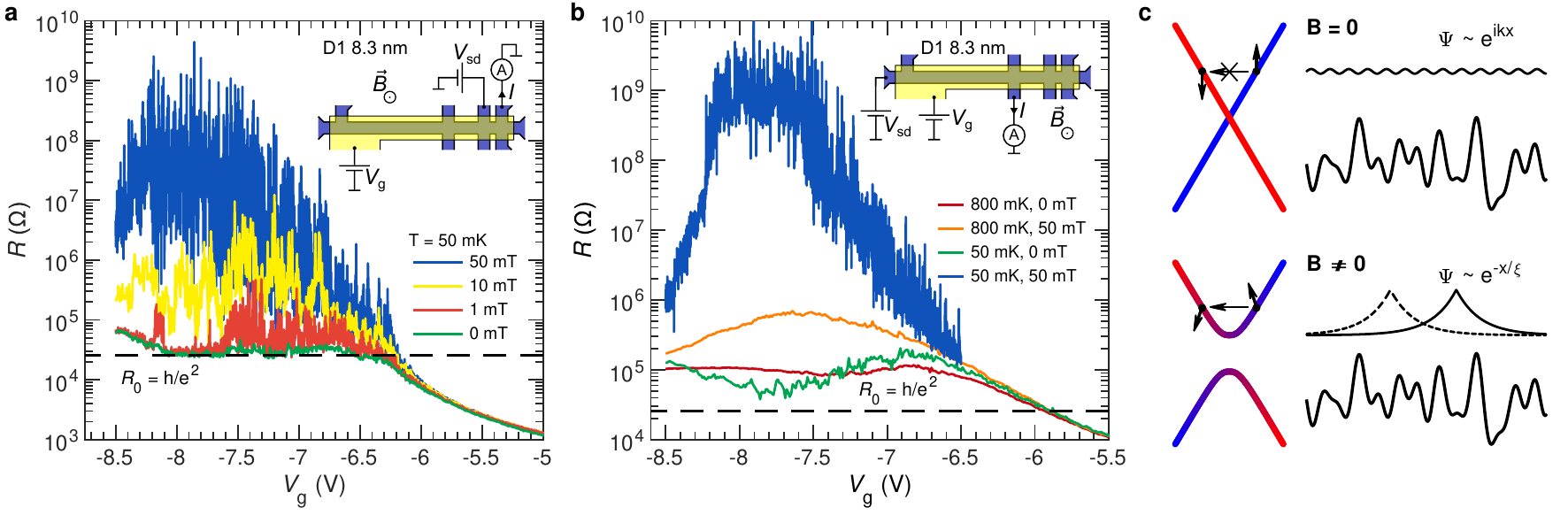}
	\end{center}
	\caption{\textbf{Two-terminal characteristics of the QSHI device.} (a)~Two-terminal linear-response resistance of the $6\,\mathrm{{\mu}m}$-long edge in~D1 ($8.3\,\mathrm{nm}$ QW) as a function of the gate voltage, measured at~$T=50\,\mathrm{mK}$ in different magnetic fields~$B_{\perp}$. (b)~Two-terminal linear-response resistance of the $38\,\mathrm{{\mu}m}$-long edge in~D1 as a function of the gate voltage, measured at~$T=50\,\mathrm{mK}$ and $T=800\,\mathrm{mK}$, with and without magnetic field~$B_{\perp}=50\,\mathrm{mT}$. The insets on (a) and (b) show device structures along with measurement schematics. (c) Effect of magnetic field on electronic states in QSHI. If~$B=0$, a gapless energy spectrum with spin-momentum locking (top left) is expected, while the propagating electrons are not influenced by potential disorder (top right). The electron wave function~$\Psi$ corresponds to free propagation along the edge~\cite{PhysRevB.93.075434}, $\Psi \sim e^{ikx}$. At non-zero~$B$ the gap opens while the counter-propagating branches are no longer comprised of states with opposite spins (bottom left). Eventually, the disorder leads to backscattering and localizes the electrons (bottom right) with their wave function envelope decaying exponentially over the localization length~$\xi$, $\Psi \sim e^{-x/\xi}$.}
	\label{hgte_mag_fig_intro}
\end{figure*}

We investigate two QSHI devices of different crystallographic orientation and QW thickness,~$d$. In device~D1, the QW with~$d=8.3\,\mathrm{nm}$ resides in a (013) plane, while the device D2 is based on the (112) QW with~$d=14\,\mathrm{nm}$. We note that the QWs of both devices are similar in design to the QWs studied in~\cite{Gusev2013,Olshanetsky2015}. Both devices are shaped as multi-terminal Hall bars with Ti/Au metallic top gates, see insets of Fig.\,\ref{hgte_mag_fig_intro}a and Fig.\,\ref{hgte_mag_fig_intro}b for the schematic representation of the device and measurement configuration. The microscope image of one of the devices may be found in Supplemental Material Fig.\,1. For additional device fabrication and measurement technique details see corresponding sections of Supplemental Material. 

Using the gate voltage~$V_{\text{g}}$, 
the Fermi level can be tuned within the bulk energy gap, as large as $30\,\mathrm{meV}$ in~D1 and $3\,\mathrm{meV}$ in~D2~\cite{Bernevig1757,Olshanetsky2015}. In this way the QSHI regime is realized, with the predominant edge conduction confirmed by transport measurements in similar devices~\cite{Gusev2011,Tikhonov2015,Olshanetsky2015,Piatrusha2017} and, independently, here via non-local resistance measurements (see Supplemental Material Fig.\,3). Various distances between the neighboring ohmic contacts allow us to choose the different lengths of the edge channels, spanning the range between $2\,\mathrm{{\mu}m}$ and $38\,\mathrm{{\mu}m}$ in each device, with edge resistance increasing with increasing edge length. The device~D1 demonstrates resistance $R \approx R_0\equiv h/e^2$ for the shortest edges, while in~D2 the resistance is about twice as large for the same edge length (see Supplemental Material Figs.\,9,\,11). In the following we discuss the results obtained for device D1. Similar data obtained for device~D2 is demonstrated in Supplemental Material.

Figs.\,\ref{hgte_mag_fig_intro}a and \ref{hgte_mag_fig_intro}b show the two-terminal resistance of the $6\,\mathrm{{\mu}m}$ and $38\,\mathrm{{\mu}m}$-long edges in~D1, with subtracted contribution of contact terminals, as a function of~$V_{\text{g}}$ in a magnetic field perpendicular to the QW plane, $B_\perp$. The measurement configurations are shown in the corresponding insets. For the two-terminal resistance of an $N$-terminal device in the phase-coherent case, Landauer-Buttiker analysis~\cite{PhysRevB.38.9375} would yield $R_0(1-1/N)$, which in our case is~$\approx0.9R_0$. This reasoning, however, is not applicable in our experiment since the edges longer than~$6\,\mathrm{\mu m}$ are in phase-incoherent regime.

For~$B_\perp=0$ at $T=50\,\mathrm{mK}$, within the range of gate voltages $-8\,\mathrm{V}<V_{\text{g}}<-6.5\,\mathrm{V}$ the $6\,\mathrm{{\mu}m}$-long edge demonstrates the conductance plateau with the value close to~$G_0$ and the device demonstrates the non-local resistance (see Supplemental Material Fig.\,3), which corresponds to the onset of the QSHI regime. At the same time, no sizable $T$-dependence is observed, which is usual for the QSHI edges~\cite{Nowack2013,Gusev2014,Tikhonov2015}. By contrast, in a small magnetic field of~$B_\perp=50$\,mT the resistance increases dramatically, sometimes reaching $R\sim1{\rm G\Omega}$ in $B_\perp=50$\,mT at~$T=50\,\mathrm{mK}$. Qualitatively similar but even stronger effect of the magnetic field is observed in the longer $38\,\mathrm{{\mu}m}$ edge, as shown in Fig.\,\ref{hgte_mag_fig_intro}b. As the temperature is raised to~$T=800\,\mathrm{mK}$, the resistance drops down again by more than a factor of~$10^3$. The straightforward crosscheck demonstrates that in all our measurements the current flows along the edges of the device, while the bulk conduction contribution remains negligible even for $R\sim1{\rm G\Omega}$ (see Supplemental Material Fig.\,4). All the edges of both our devices~D1 and D2 exhibit the reported resistance increase in a small magnetic field which is also the case for the four-terminal configuration measurements (see Supplemental Material Figs.\,9--12 for additional data).

Figs.\,\ref{hgte_mag_fig_intro}a and\,\ref{hgte_mag_fig_intro}b highlight our main result that a tiny magnetic field gives rise to the dramatic increase of the resistance of the helical edge states accompanied by strong $T$-dependence and highly reproducible, giant mesoscopic fluctuations (see Supplemental Material Fig.~5). Altogether, this behavior is a hallmark of the Anderson localization of the electronic states and manifests a transition from the topologically protected phase to the trivial insulator in a magnetic field as anticipated in various scenarios~\cite{PhysRevB.93.075434,PhysRevLett.109.246803,Raichev2015}. The underlying microscopic explanation is depicted in Fig.\,\ref{hgte_mag_fig_intro}c. In~$B=0$, the dispersion relation of the helical electrons consists of two opposite spin counter-propagating branches, the coherent backscattering between which is forbidden by the topological protection~\cite{RevModPhys.83.1057}. In a finite $B$-field, the branches hybridize, opening an energy gap in the vicinity of the Dirac point, and the spins of the counter-propagating electrons acquire a common tilt along the magnetic field. Thus, the coherent backscattering mediated by potential disorder is restored in the magnetic field, resulting in Anderson localization of the electronic states on the length-scale of the mean-free path, in contrast to the $B=0$~case.

It is important to emphasize that both short edges with $G\approx G_0$ and the longer ones with $G<G_0$ behave similarly in finite~$B$. Breaking the TRS with magnetic field allows coherent backscattering in both cases, leading to the helical edge states localization. In this scenario, the observation of $G<G_0$ for long edges in TRS-case is due to phase-breaking of the origin yet to be explained~\cite{PhysRevX.3.021003,Ma2015,Xu2006,PhysRevLett.102.256803,Vayrynen2013,Kainaris2014,Wang2017,Vayrynen2018,PhysRevLett.122.016601}, while the absence of localization in zero magnetic field for resistive edges manifests the topological protection.
	
\begin{figure}
	\begin{center}
		\includegraphics[scale=1]{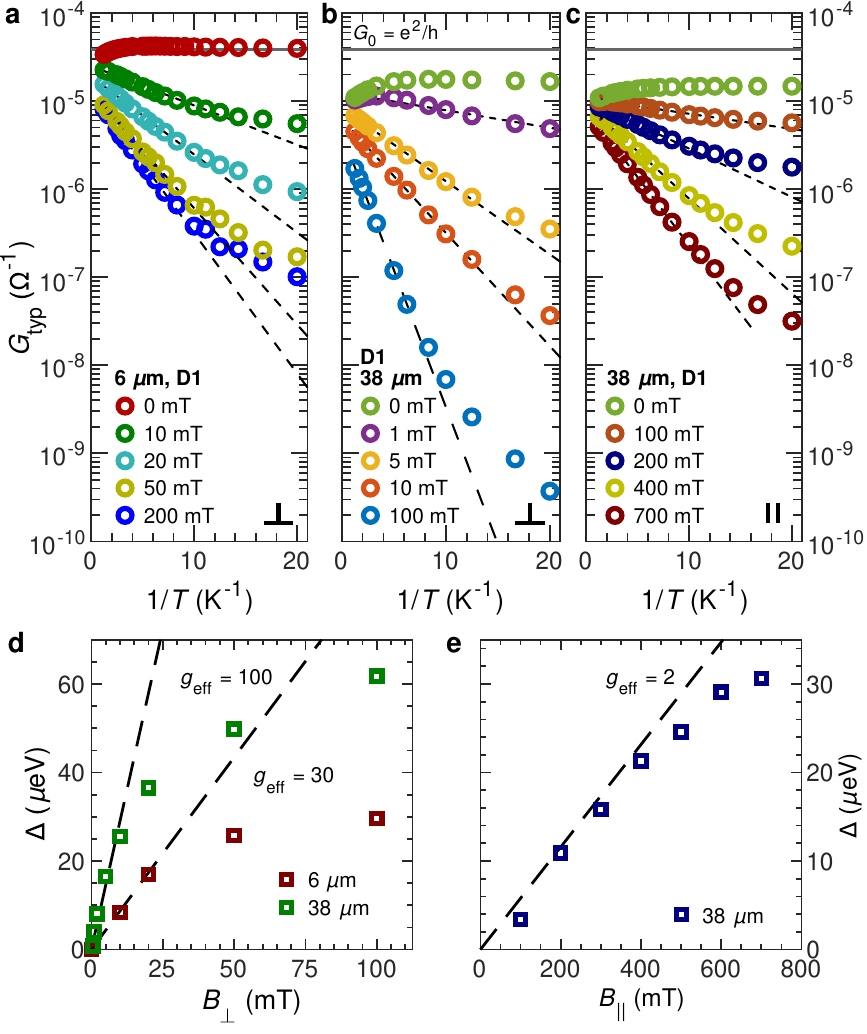}
	\end{center}
	\caption{\textbf{Temperature and magnetic field dependence of the edge conductance in the localized regime.} Log-averaged conductance~$G_\mathrm{typ}$ (see text) as a function of inverse temperature of the (a)~$6\,\mathrm{{\mu}m}$-long edge in~D1 and of the (b,c)~$38\,\mathrm{{\mu}m}$-long edge in $B_\perp$ and~$B_\parallel$, oriented at~$45^\circ$ to the edge, respectively. Application of magnetic field leads to the activation-like $G_\mathrm{typ}(T)$, in contrast to the weak metallic $G_{\mathrm{typ}}(T)$ in~$B=0$. The strong anisotropy with respect to the $B$-orientation is evident. Solid lines demonstrate the conductance quantum value~$G_0$. Panels~(a-c) share the same y-axis. Slight difference between zero-field data in panels~(b,c) is due to thermal recycling. The example of raw data is presented in Supplemental Material Fig.~7. (d)~Activation energy~$\Delta$ (see text) as a function of~$B_{\perp}$, extracted from measurements in (a,b). The dashed lines demonstrate the magnitude of the effect and are $2\Delta = g_{\text{eff}} \mu_\mathrm{B} B_{\perp}$ with the specified values of~$g_{\text{eff}}$. (e)~The same as~(d), but for the~$38\,\mathrm{{\mu}m}$-long edge in~$B_\parallel$, extracted from measurements in~(c).}
	\label{hgte_mag_fig_T}
\end{figure}

We now discuss in detail the $T$-dependence of the QSH edge conductance $G = 1/R$ under broken TRS. As seen from Fig.\,\ref{hgte_mag_fig_T}, in the absence of magnetic field the $T$-dependence within the charge neutrality point (CNP) region is of metallic type with approximately $20\%$ change of the $6\,\mathrm{{\mu}m}$-long edge resistance as $T$ is reduced from $800\,\mathrm{mK}$ to $50\,\mathrm{mK}$. For the $38\,\mathrm{{\mu}m}$-long edge, 
within the range of gate voltages $-8.3\,\mathrm{V}<V_{\text{g}}<-7.4\,\mathrm{V}$ we observe even stronger metallic behavior with a two fold decrease of resistance in the same $T$ interval. We are not aware of similar observations in HgTe QWs, where the reported $T$-dependencies are usually either completely absent or weakly insulating. We note, however, that most of the studies discuss $R(V_{\text{g}})$-dependencies at temperatures above~$1\,\mathrm{K}$. Additionally, we note that in~D2, as well as for the $38\,\mathrm{{\mu}m}$-long edge in~D1 within the range of gate voltages $-7.4\,\mathrm{V}<V_{\text{g}}<-6.5\,\mathrm{V}$, the $R(T)$-dependence in~$B=0$ is of weakly insulating type at any~$V_{\mathrm{g}}$ (see Supplemental Material Fig.\,6).

The $G(T)$-dependence in a magnetic field is much more impressive. In the presence of strong fluctuations, we analyze the log-averaged (typical)~\cite{RevModPhys.69.731} conductance 
$G_{\text{typ}}\equiv G_\mathrm{0}\exp\left\langle \ln {G}/{G_\mathrm{0}} \right\rangle$, 
with the averaging performed over the small gate-voltage region $-7.65\,\mathrm{V}<V_{\text{g}}<-7.55\,\mathrm{V}$ within the resistance maximum in Fig.\,\ref{hgte_mag_fig_intro}b. This relatively narrow range of~$V_{\text{g}}$ is chosen in order to resolve the fine structure of mesoscopic fluctuations in the vicinity of $V_{\text{g}}$-value where the $G(T)$-dependence is prominent enough, as reflected by the gate voltage dependence of the activation energy (see Supplemental Material Fig.\,8). The resulting data is shown in Fig.\,\ref{hgte_mag_fig_T}(a-c). Strikingly, already for $B_\perp\sim1$\,mT the trend of $G_\mathrm{typ}(T)$ changes from metallic to activated insulating dependence $G_{\text{typ}}\propto\,\exp(-\Delta/k_B T)$ (the remnant $B$-field did not exceed $2\,\mathrm{mT}$ and was compensated in the experiment with $0.1\,\mathrm{mT}$ precision). The activation energy reaches about $\Delta\approx10\,\mathrm{{\mu}eV}$ in $10\,\mathrm{mT}$ for the $6\,\mathrm{{\mu}m}$-long edge, shows a sub-linear increase with $B_\perp$ and increases with the length of the edge, Fig.\,\ref{hgte_mag_fig_T}d. Thus, it is difficult to make an obvious relation of the activated behavior with the single-particle spectrum of the helical edge states, e.g. with a Zeeman gap opening at the Dirac point. Similar observations hold for the in-plane orientation of the magnetic field, see Figs.\,\ref{hgte_mag_fig_T}(c,e) for the case of $B_{\parallel}$ directed at about $45^\circ$ with respect to the edge under study. Here, the magnetic fields~$B_\parallel$ roughly an order of magnitude stronger are required to observe the activated behavior comparable to the $B_\perp$ case. This might be a consequence of the Lande $g$-factor anisotropy predicted for HgTe QWs in some works~\cite{PhysRevB.93.075434}.
	
\begin{figure}
	\begin{center}
		\includegraphics[scale=1]{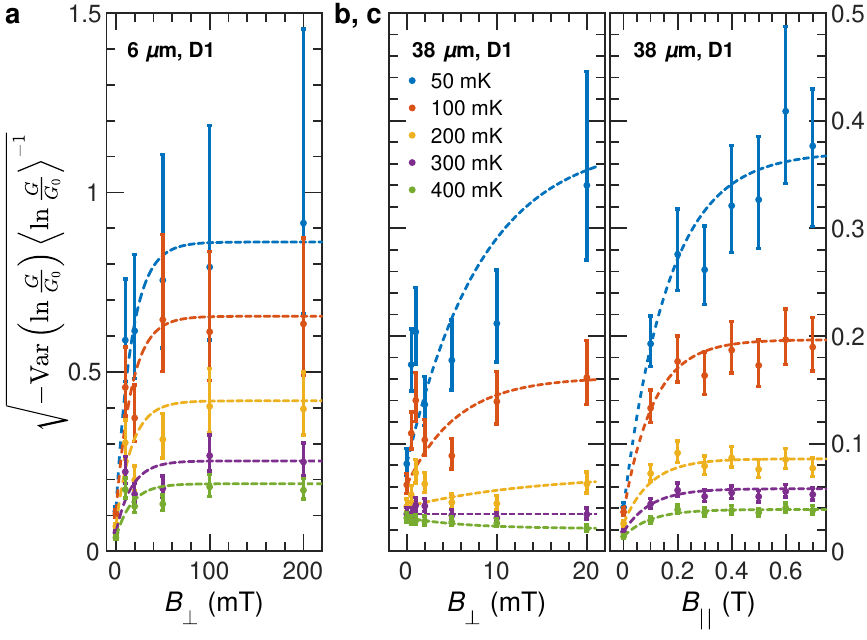}
	\end{center}
	\caption{\textbf{Conductance fluctuations in the localized regime and it's temperature dependence}. Square root of the normalized variance of the logarithm of the edge conductance (see text) of the (a)~$6\,\mathrm{{\mu}m}$-long edge in~D1 as a function of~$B_\perp$ and of the (b,c)~$38\,\mathrm{{\mu}m}$-long edge as a function of~$B_\perp$ and $B_\parallel$. The confidence intervals are due to statistical analysis assuming the log-normal distribution of the conductance. The peak at lowest magnetic fields in panel~(b) is most likely due to the sudden change of the state of the scatterers for a given edge. Dashed lines are guides for an eye. Panels~(b) and~(c) share the same y-axis.}
	\label{hgte_mag_fig_Var}
\end{figure}

\begin{figure}
	\begin{center}
		\includegraphics[scale=1]{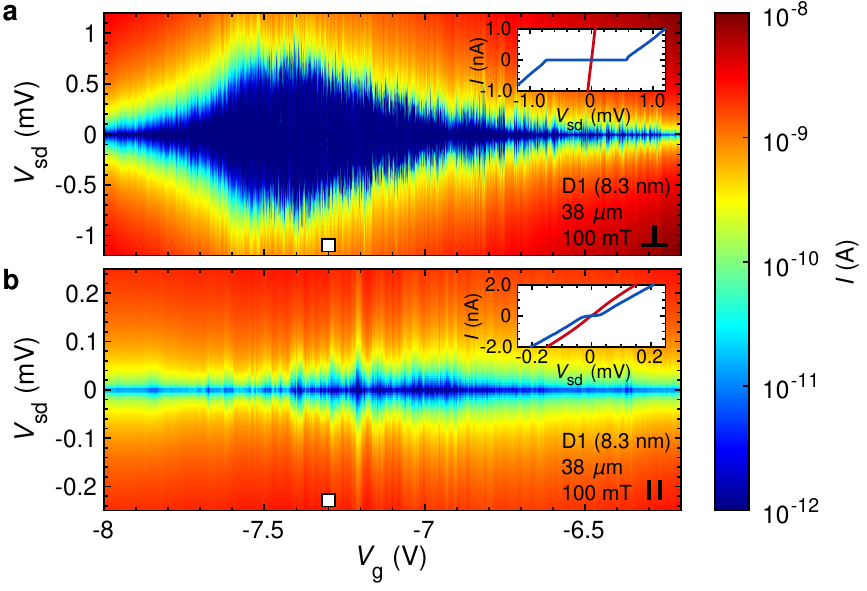}
	\end{center}
	\caption{\textbf{Gap opening in the edge by external magnetic field.} (a)~Current, flowing through the~$38\,\mathrm{{\mu}m}$-long edge of D1 as a function of gate and bias voltages in $B_{\perp}=100\,\mathrm{mT}$. Measured current is plotted as a logarithm of its absolute value (see text). The inset demonstrates $I-V_\mathrm{sd}$ curves, measured at $V_{\text{g}}=-7.3\,\mathrm{V}$ (white square) in $B_{\perp}=0$ (red line) and $100\,\mathrm{mT}$ (blue line). (b)~The same as~(a) for $B_{\parallel}=100\,\mathrm{mT}$, oriented at $45^\circ$ to the edge.}
	\label{hgte_mag_fig_gaps}
\end{figure}

We now quantitatively analyze the observed giant conductance fluctuations, which are the distinctive feature of the Anderson localized phase.
Here, in contrast to the metallic phase, the conductance is exponentially sensitive to the minor variations of disorder, or equivalently, to the Fermi energy. As a result, the fluctuations of the conductance are as large as the average value and obey the log-normal distribution, i.e. it is the quantity 
$\ln G/G_0$ 
which is Gaussian-distributed~\cite{RevModPhys.69.731}. In Fig.\,\ref{hgte_mag_fig_Var} we study the conductance fluctuations as a function of magnetic field at various~$T$. The square root of the normalized variance of the logarithm of the conductance, given by $\sqrt{-\text{Var}\left(\ln{G}/{G_0}\right)\left\langle\ln{G}/{G_0}\right\rangle^{-1}}$,
is plotted in panels (a,b) for short and resistive edges in $B_{\perp}$ and in panel (c) for resistive edge in $B_{\parallel}$. At the lowest available $T=50\,\mathrm{mK}$ it increases by more than one order of magnitude, compared to the $B=0$ case, for $B_\perp\approx20\,\mathrm{mT}$ and $B_\parallel\approx500\,\mathrm{mT}$. Still, the observed values are below the theoretical value of $\sqrt2$, which would correspond to a quantum-coherent Anderson localized phase~\cite{RevModPhys.69.731}. We attribute the difference to the impact of averaging in the presence of dephasing~\cite{Xu2006,PhysRevLett.102.256803,Vayrynen2013,Kainaris2014,Wang2017,Vayrynen2018,PhysRevLett.122.016601}, which also qualitatively explains the strong $T$-dependence of the fluctuations in Fig.\,\ref{hgte_mag_fig_Var}. The stronger fluctuations in the shorter edge further support this suggestion.

To further characterize the transport properties of the localized states, we study the non-linear transport regime originating from the delocalization of the edge states by the electric field~\cite{PhysRevB.46.13303,PhysRevB.58.8009} for the $38\,\mathrm{{\mu}m}$-long edge in~D1 at $T=50\,\mathrm{mK}$. The dependence of $I$ on $V_{\mathrm{sd}}$ and $V_{\text{g}}$ is plotted in panels (a) and (b) of Fig.~\ref{hgte_mag_fig_gaps} for $B_{\perp}=100\,\mathrm{mT}$ and $B_{\parallel}=100\,\mathrm{mT}$, respectively. Within the CNP region one can see the dramatic changes in $I$ at increasing $|V_{\mathrm{sd}}|$ (note the log-scale). Below the certain threshold value of $V_{\mathrm{sd}}$, which depends on $V_{\text{g}}$ with pronounced reproducible fluctuations above that, we observe only the negligible current $|I|<1\,\mathrm{pA}$ through the edge. For the magnetic field of $100\,\mathrm{mT}$, the typical bias range of suppressed conduction changes from about $0.5\,\mathrm{mV}$ in $B_\perp$ to about $50\,\mathrm{{\mu}V}$ in~$B_\parallel$. For this threshold, the corresponding energy scale is considerably higher than the activation energy extracted from the $T$-dependencies similar to that of Fig.~\ref{hgte_mag_fig_T}, indicating that the applied bias is shared among a few strongly localized electronic states along the edge. For the case of metal-insulator transition in Si inversion layers similar reasoning was suggested in~\cite{PhysRevB.46.13303}. Above the threshold, the conduction reasonably comparable to the $B=0$ case is restored. Two representative $I$-$V_{\mathrm{sd}}$ cuts of Figs.~\ref{hgte_mag_fig_gaps}a and~\ref{hgte_mag_fig_gaps}b at $V_{\text{g}}=-7.3\,\mathrm{V}$ are detailed in the corresponding insets, along with the $I$-$V_{\mathrm{sd}}$ curves at $V_{\text{g}}=0$. The observed highly non-linear transport behavior is yet another evidence of the Anderson localization of the helical states driven by the $B$-field and contrasts with the almost linear current-voltage response in the TRS $B=0$ case (see Supplemental Material Figs.\,4,\,9--11).

In conclusion, through the low-temperature magnetoresistance measurements we were able to directly demonstrate the actual strength of topological protection in one-dimensional helical edge states of HgTe-based topological insulators. Breaking the TRS with an external magnetic field allowed us to expose the hallmark Anderson localization features for the edge states – the exponential $T$-dependence, giant reproducible mesoscopic fluctuations and the gap-opening-like features in the $I$-$V_\mathrm{sd}$ characteristics. Our observation of identical behavior of the edges independently of their resistance in the finite~$B$, strongly suggests the loss of topological protection at broken TRS. At the same time, the tremendous $G(T)$-dependence of the edges in~$B\neq0$, almost absent in zero~$B$, highlights the distinct action of TRS breaking compared to other mechanisms, which enable backscattering in~$B=0$. The observation of ballistic transport via helical edge states strongly depends on the material, it’s quality, may be even sample dependent and presently does not go beyond edges several micrometers long. Our experiment demonstrates that for resistive edges the topological protection is still there in the time-reversal symmetric case perfectly sustaining the edge transport from localization.

All the measurements were performed under the support of Russian Science Foundation Grant No. 18-72-10135. The high-resistance measurements were developed and tested within the state task of ISSP RAS. We thank  I.V.\,Gornyi, T.M.\,Klapwijk, A.D.\,Mirlin, D.V.\,Shovkun and S.A.\,Tarasenko for helpful discussions.


%

\widetext
\clearpage
\begin{center}
\textbf{\large Supplemental Material}
\end{center}
\setcounter{equation}{0}
\setcounter{figure}{0}
\setcounter{table}{0}
\setcounter{page}{1}
\makeatletter
\renewcommand{\theequation}{S\arabic{equation}}
\renewcommand{\figurename}{Supplemental Material Fig.}
\renewcommand{\bibnumfmt}[1]{[S#1]}
\renewcommand{\citenumfont}[1]{S#1}

\section*{Device fabrication}
The single HgTe QWs as heterojunction $\mathrm{Cd}_{0.7}\mathrm{Hg}_{0.3}\mathrm{Te}/\mathrm{HgTe}/\mathrm{Cd}_{0.7}\mathrm{Hg}_{0.3}\mathrm{Te}$ were grown by molecular beam epitaxy on SI GaAs substrates with buffer ZnTe and CdTe layers CdTe with the thickness of $0.1\,\mathrm{\mu m}$ and $5$-$7\,\mathrm{\mu m}$, respectively. The mesa was fabricated via plasma chemical
etching followed by covering with $\text{SiO}_2/\text{Si}_3\text{N}_4$ insulating layers, $200\,\mathrm{nm}$ thick in total. All devices are equipped with evaporated Au/Ti metallic gates.
\begin{figure*}[h]
	\begin{center}
		\includegraphics[scale=1]{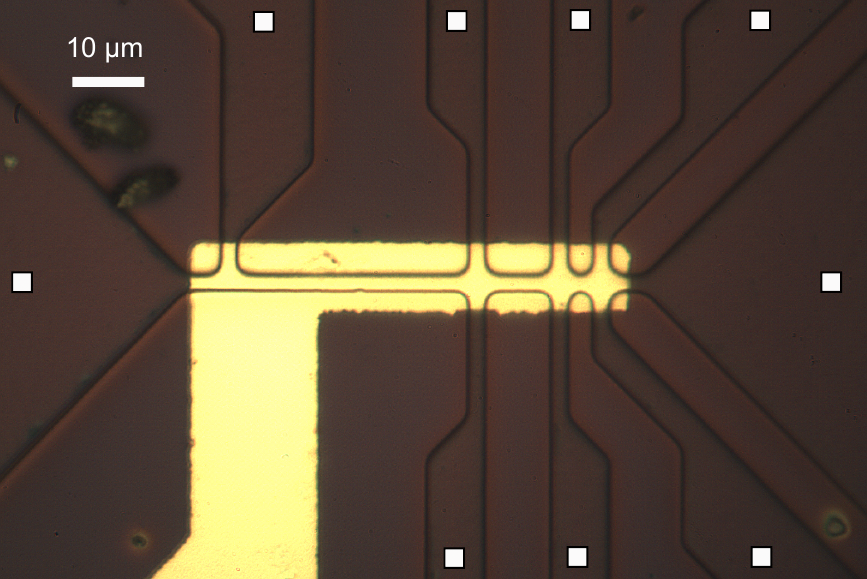}
	\end{center}
	\caption{\textbf{Optical microphotograph of the 14\,nm HgTe/CdHgTe device D2.} The regions marked with white dots correspond to the contact leads with the hall bar in between, while in the other regions the QW is completely etched. The central hall-bar region is covered with the Ti/Au top gate (yellow).}
	\label{hgte_supp_sample}
\end{figure*} 

\newpage
\section*{Measurement techniques} 
All devices were measured in a BlueFors-LD$250$ dilution refrigerator with a base temperature of $17\,\mathrm{mK}$ equipped with a $9\,\mathrm{T}$ superconducting solenoid. The lowest obtainable electronic temperature was $T \approx 50\,\mathrm{mK}$, 
{verified by the Johnson-Nyquist thermometry}. 
The two-terminal linear response resistances were obtained by differentiating of the $I$-$V_{\text{sd}}$ curves, measured via the transimpedance amplifier with $1\,\mathrm{G\Omega}$ coefficient -- see Supplemental Material Fig.~2. For all the data we subtract the contact resistance which was determined separately and typically ranged between~$2$~and $5\,\mathrm{k\Omega}$  for each terminal in our devices. Presenting the data, we also do not take into account the parallel conduction channel which appears in the QSHI regime and which is due to transport in the counterdirection for a given edge. This parallel channel would lead to the correction of approximately $30\,\%$ for the $38\,\mathrm{\mu m}$-long edge in~D1 and of approximately $10\,\%$ for the $12\,\mathrm{\mu m}$-long edge in~D2 in zero magnetic field. It is harder to estimate the correction in the finite~$B$ due to the absence of self-averaging of resistance of the edge channels in the insulating regime.

\begin{figure*}[h]
	\begin{center}
		\includegraphics[scale=1]{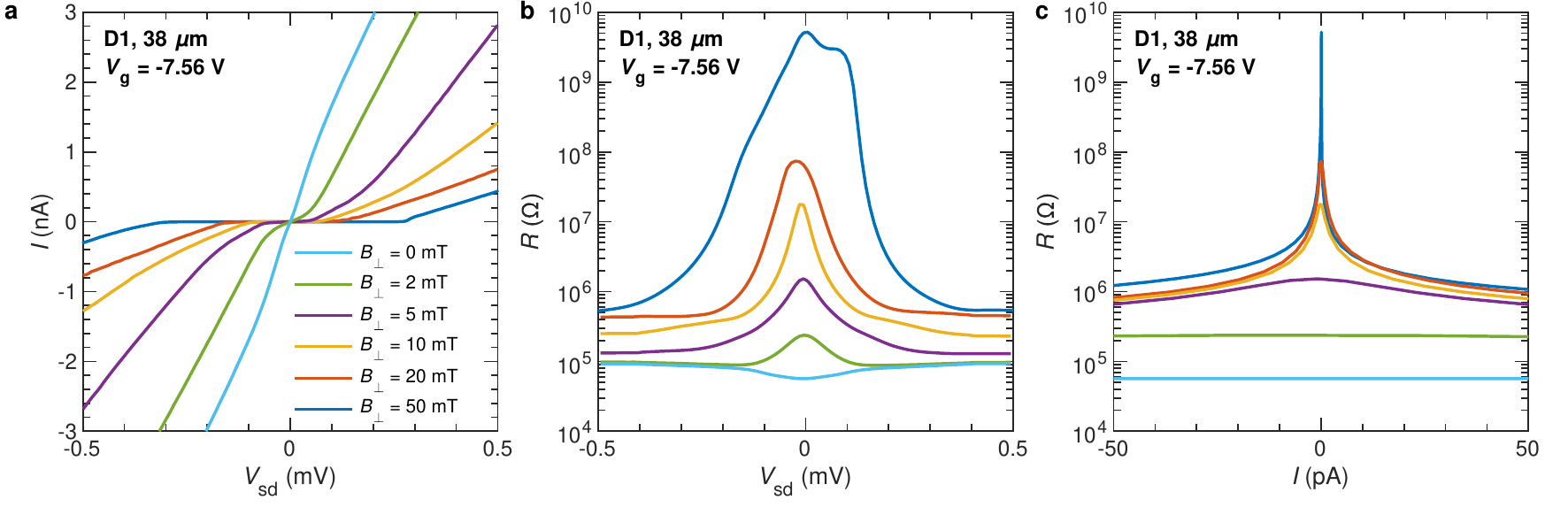}
	\end{center}
	\caption{\textbf{Linear-response resistance in the localized regime.} (a)~$I$-$V_{\text{sd}}$ curves of the $38\,\mathrm{\mu m}$-long edge measured at $V_{\text{g}}=-7.56\,\mathrm{V}$, corresponding to the QSHI regime in $B=0$. (b,c)~Differential resistance $R=1/\left(dI/dV_{\mathrm{sd}}\right)$ is obtained by the differentiation of the $I$-$V_{\text{sd}}$ curves and is plotted as a function of~$V_{\text{sd}}$ and~$I$. In studying the linear-response resistance in figures~1--3 of the main text we take thus obtained values $R=R(V_{\text{sd}}=0)$.}
	\label{hgte_supp_diff}
\end{figure*}

\newpage
\begin{figure*}[h]
	\begin{center}
		\includegraphics[scale=1]{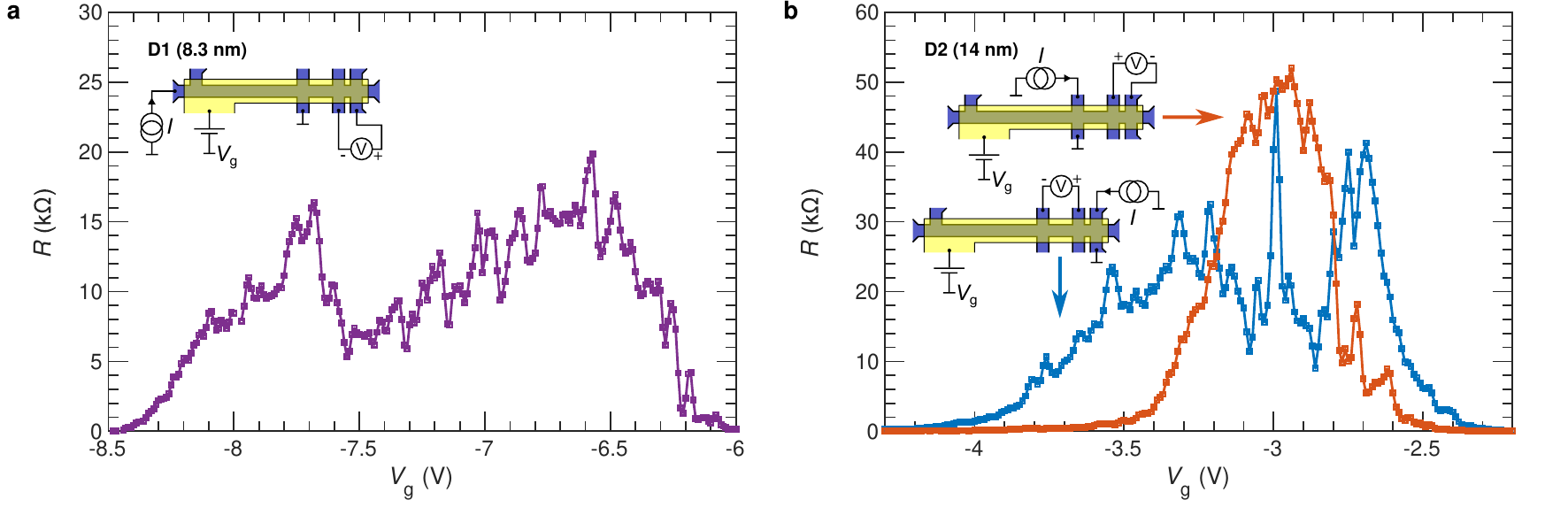}
	\end{center}
	\caption{\textbf{Non-local resistance measurements in devices D1 and D2.} Non-local resistance as a function of gate voltage, measured via conventional lock-in technique in the configurations indicated in the insets for devices (a) D1 and (b) D2. The data was taken at $T=50\,\mathrm{mK}$. Note, that the $V_{\text{g}}$-region where non-local signal is observed may differ for different edges and may float by several hundred millivolts with thermal recycling or after long staying in large negative $V_{\text{g}}$. This explains why the data of Figs.\,1, 4 and Supplemental Material Figs.\,3, 7 and 9-11 may differ in terms of $V_{\text{g}}$-range where the devices demonstrate the QSHI behavior.}
	\label{hgte_supp_nonlocal}
\end{figure*}

\newpage
\begin{figure*}[h]
	\begin{center}
		\includegraphics[scale=1]{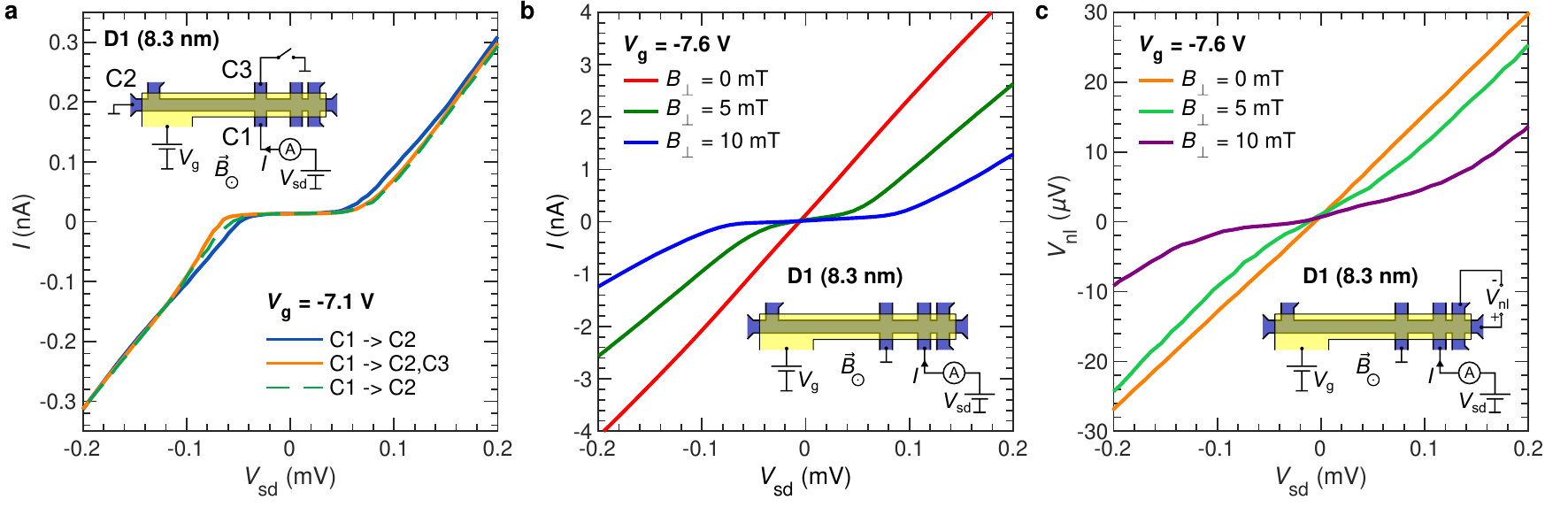}
	\end{center}
	\caption{\textbf{Edge transport in localized mode in~D1.} (a) Two-terminal conductance measurements in $B_{\perp}=50$\,mT at $T=50\,\mathrm{mK}$ with two different contact configurations (see inset schematic). Bias voltage $V_\mathrm{sd}$ is applied to the contact C1, where the total flowing current $I$ is measured. The solid blue line corresponds to the configuration, when only C2, positioned at $38\,\mathrm{{\mu}m}$ from C1, is grounded. The solid orange line corresponds to the situation, when C3 is additionally grounded. The dashed green curve is the repeated measurement of the blue curve directly after the orange one. The insignificant difference between all three curves verifies the edge transport domination. (b) Two-terminal conductance measurements in different magnetic fields for the configuration indicated in the inset. (c) Non-local voltage $V_{\mathrm{nl}}$, measured in different magnetic fields for the configuration indicated in the inset. The presence of finite slope of $V_{\text{nl}}$ inside the transport gap indicates the edge transport presence.}
	\label{hgte_mag_edge_bulk}
\end{figure*}

\newpage
\begin{figure*}[h]
	\begin{center}
		\includegraphics[scale=1]{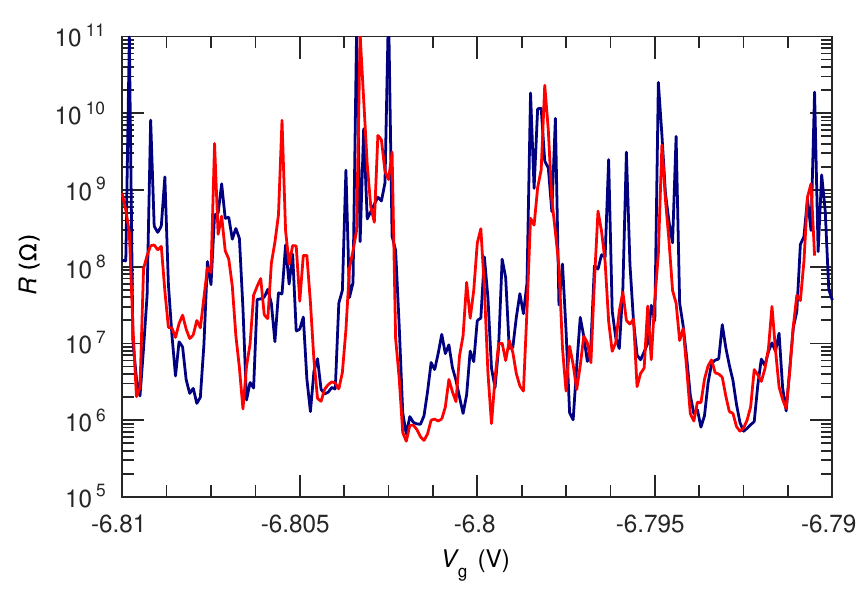}
	\end{center}
	\caption{\textbf{Mesoscopic fluctuations of the edge resistance.} The fluctuations in the $R(V_{\text{g}})$-dependence of the $38\,\mathrm{\mu m}$-long edge in D1 at $T=50\,\mathrm{mK}$ in $B_{\perp}=100\,\mathrm{mT}$. Each data point here is obtained by differentiating of the corresponding $I-V_{\mathrm{sd}}$ curve, as shown in Supplemental Material Fig.~2. The fluctuations in the two consecutive sweeps are highly reproducible verifying its mesoscopic origin. Sweeping in a significantly larger $V_{\text{g}}$-range probably leads to charge retrapping in the gate insulator and the fluctuations stop being perfectly reproducible. Note also, that the four orders of magnitude fluctuations observed here are observed in Supplemental Material Fig.\,9(d) at slightly different values of gate voltages near $V_{\text{g}}=-7\,\mathrm{V}$, which is due to the long term $V_{\text{g}}$-drift by approximately $0.2\,\mathrm{V}$.}
	\label{hgte_additional_14nm}
\end{figure*}	

\newpage
\begin{figure*}[h]
	\begin{center}
		\includegraphics[scale=1]{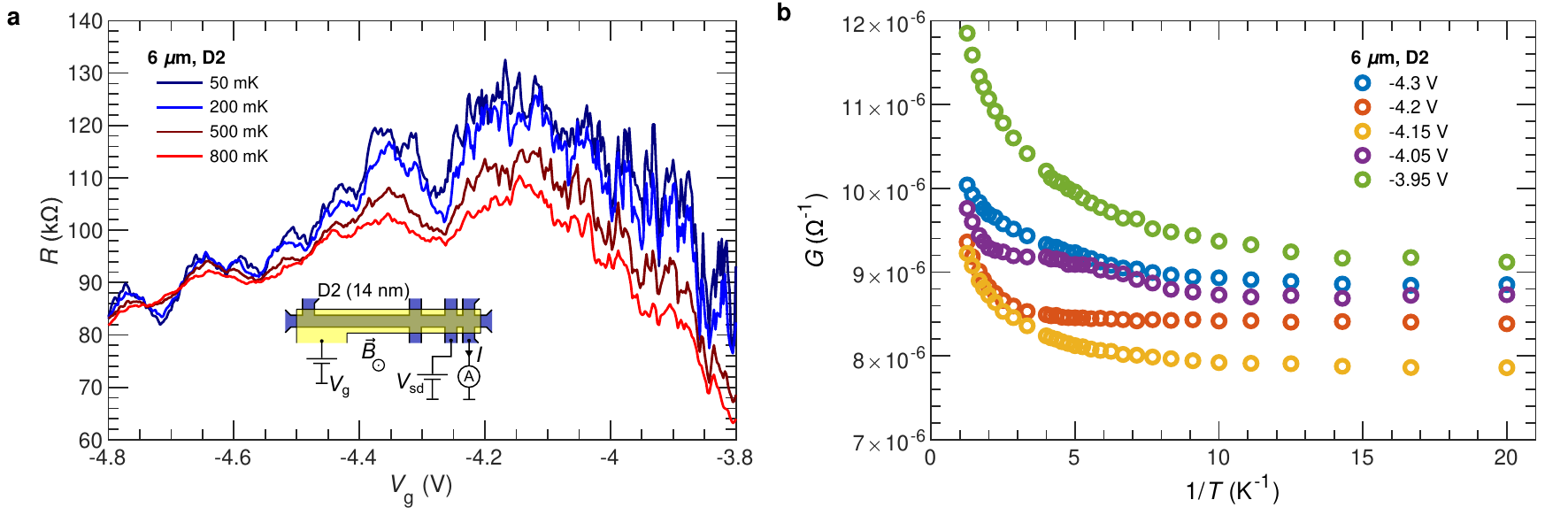}
	\end{center}
	\caption{\textbf{Temperature dependence of the $\mathbf{6\,\boldsymbol{\mu}m}$-long edge resistance in device D2 in zero magnetic field}. (a) Edge resistance as a function of the gate voltage $V_\mathrm{g}$ for four $T$ values. The measuruments were performed in the configuration of the inset. (b) Conductance of the same edge, as in (a) versus $1/T$ for fixed gate voltages (see legend).}
	\label{hgte_14nm_R_T}
\end{figure*}

\newpage
\begin{figure*}[h]
	\begin{center}
		\includegraphics[scale=1]{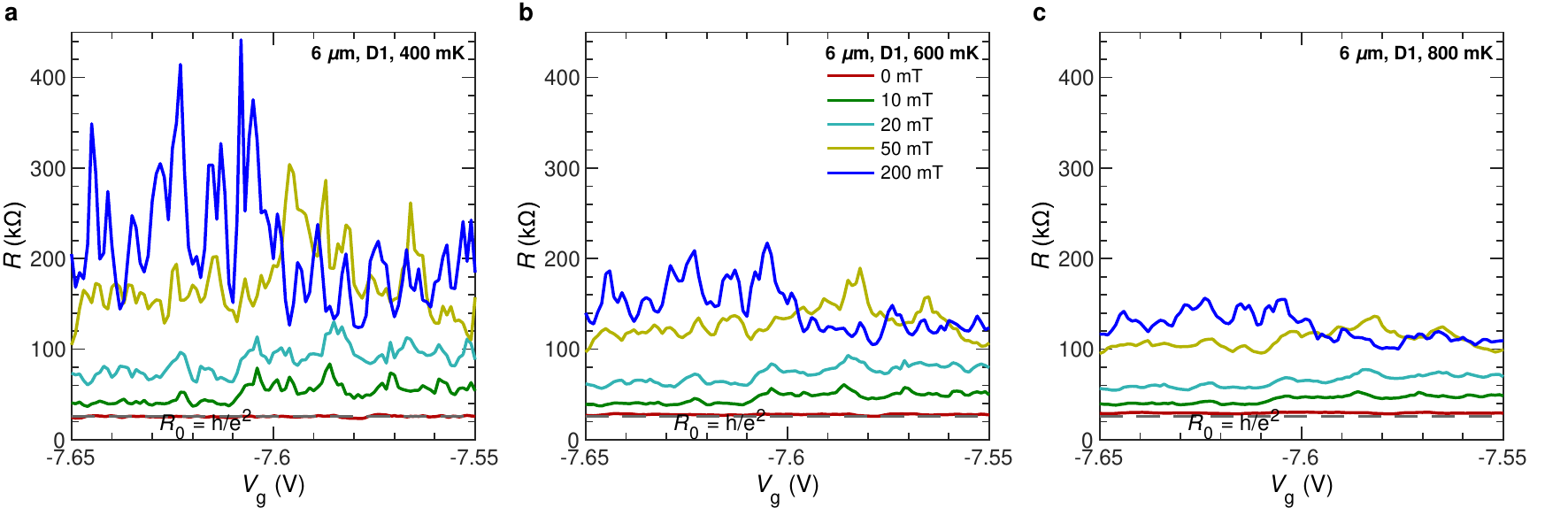}
	\end{center}
	\caption{\textbf{Temperature and magnetic field evolution of the edge resistance.} $R(V_{\text{g}})$-dependencies of the $6\,\mathrm{{\mu}m}$-long edge in~D1 at various $B_{\perp}$ at (a)~$T=400\,\mathrm{mK}$; (b)~$T=600\,\mathrm{mK}$; and (c)~$T=800\,\mathrm{mK}$.}
	\label{hgte_additional_14nm}
\end{figure*}

\newpage
\begin{figure*}[h]
	\begin{center}
		\includegraphics[scale=1]{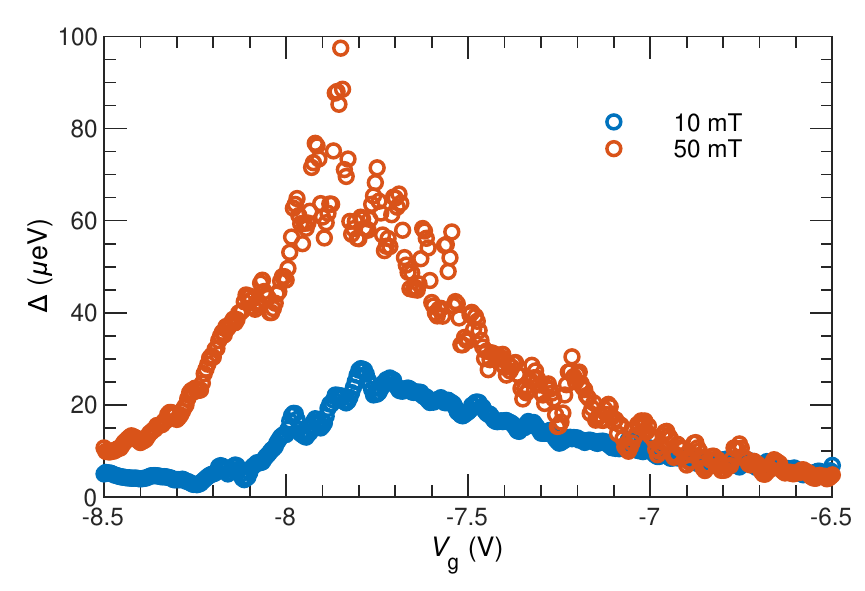}
	\end{center}
	\caption{\textbf{Rough estimate for the activation energy $\Delta$ in finite~$B$ in the QSHI regime.} The data for the $38\,\mathrm{\mu m}$-long edge in D1 is obtained by analyzing the $G(T)$-dependence in the range between $80\,\mathrm{mK}$ and $0.8\,\mathrm{K}$. For the thorough analysis, in the main text we concentrate on the small gate-voltage region $-7.65\,\mathrm{V}<V_{\text{g}}<-7.55\,\mathrm{V}$ where the $G(T)$-dependence is prominent.}
	\label{hgte_additional_14nm}
\end{figure*}


\newpage
\begin{figure*}[h]
	\begin{center}
		\includegraphics[scale=1]{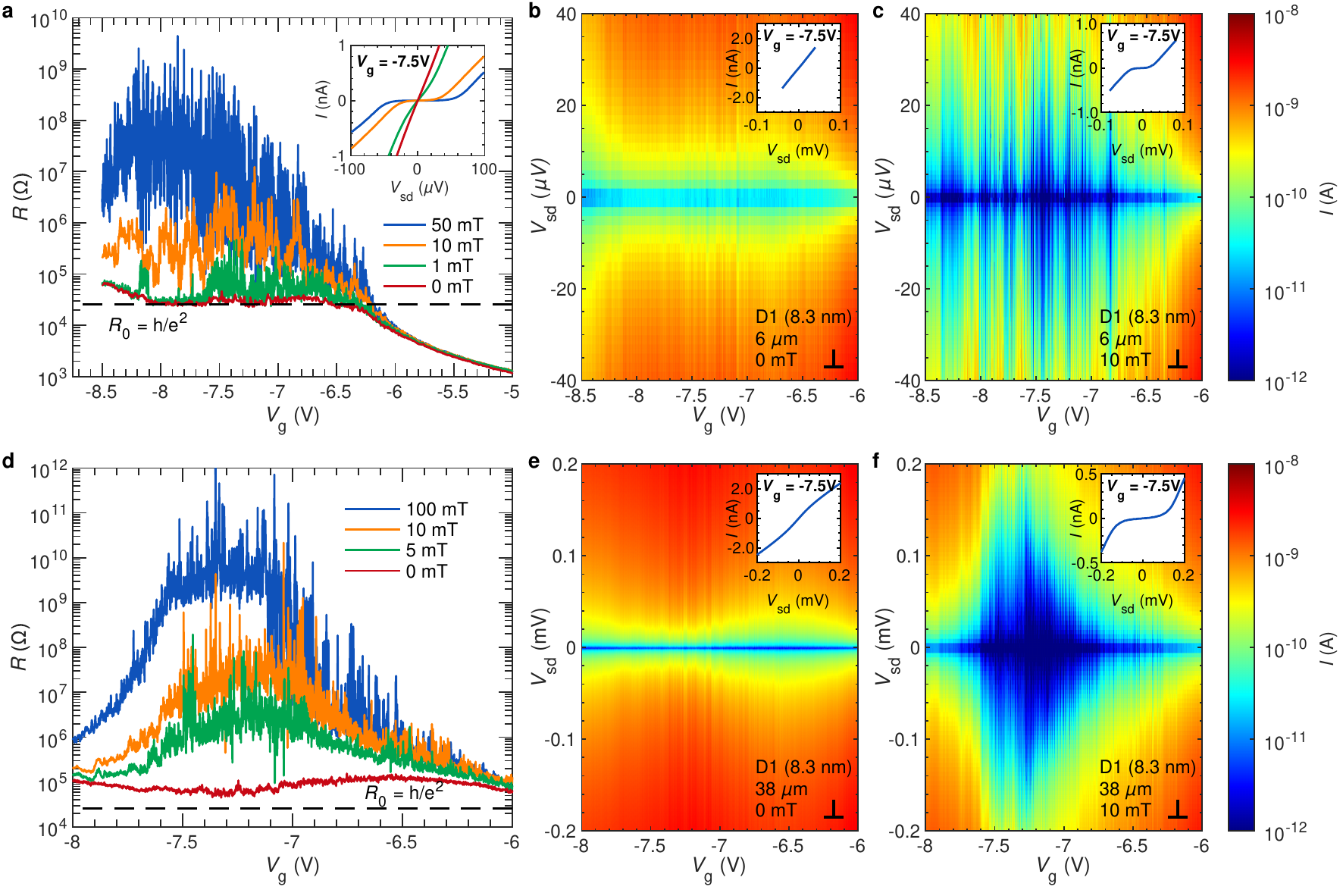}
	\end{center}
	\caption{\textbf{The effect of perpendicular magnetic field in the $\mathbf{8.3\,\mathbf{nm}}$ QW.} (a) The two-terminal $R(V_{\text{g}})$-dependence of a $6\,\mathrm{{\mu}m}$-long edge in~D1 at various $B_{\perp}$ and (b,c) the gap-opening in the same edge at $B_{\perp}=10\,\mathrm{mT}$, all measured at $T=50\,\mathrm{mK}$. The insets demonstrate the corresponding $I$-$V_{\mathrm{sd}}$ curves at the specified $V_{\text{g}}$ values. (d-f) The same measurements for the $38\,\mathrm{{\mu}m}$-long edge. Note that the data was taken in another cooling compared to Fig.\,1a hence a horizontal shift of $R(V_{\mathrm{g}})$-dependence.}
	\label{hgte_additional_8nm}
\end{figure*}

\newpage
\begin{figure*}[h]
	\begin{center}
		\includegraphics[scale=1]{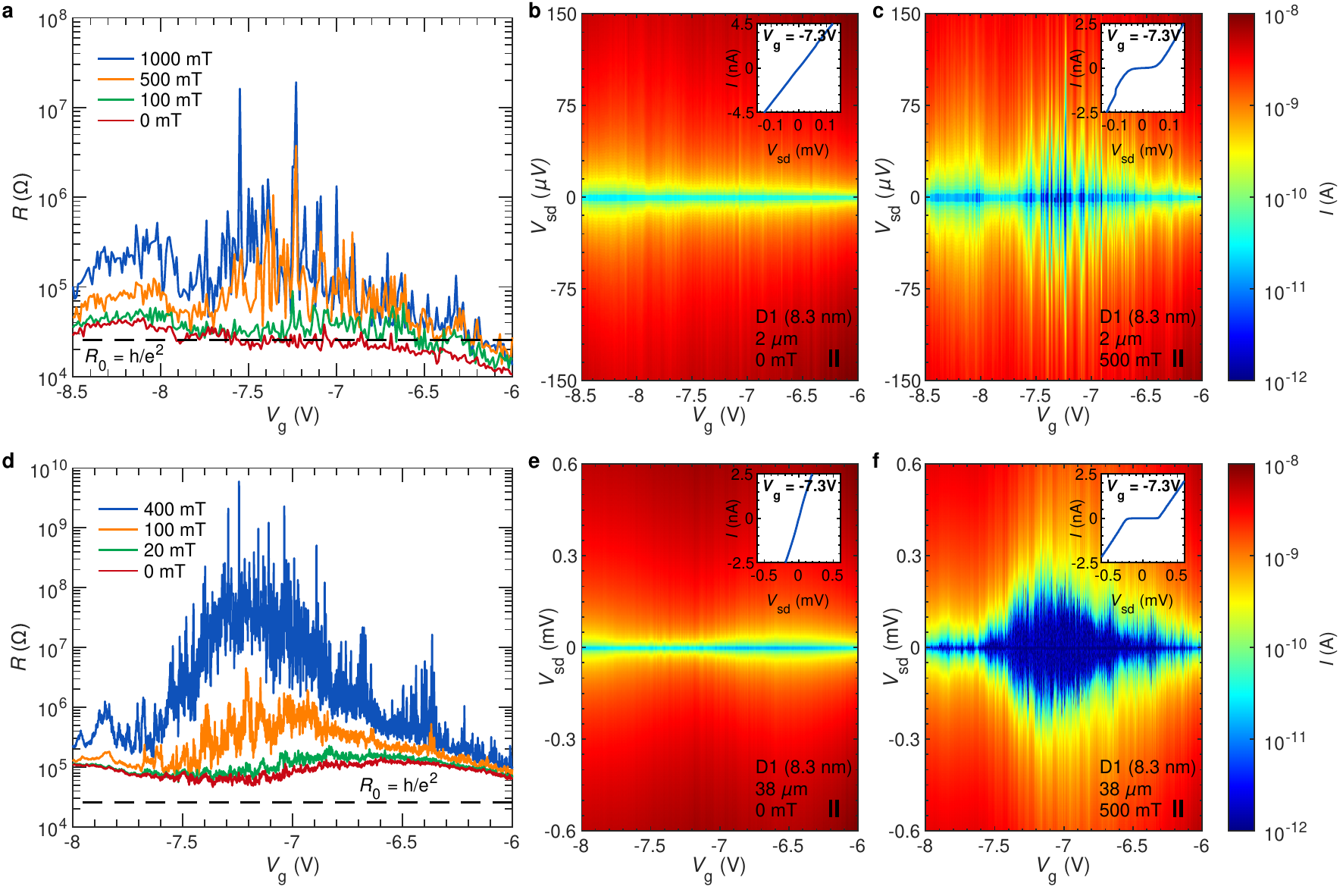}
	\end{center}
	\caption{\textbf{The effect of in-plane magnetic field in the $\mathbf{8.3\,\mathbf{nm}}$ QW.} (a) The two-terminal $R(V_\text{g})$-dependence of a $2\,\mathrm{{\mu}m}$-long edge in~D1 at various $B_{\parallel}$ and (b,c) the gap-opening in the same edge at $B_{\parallel}=500\,\mathrm{mT}$, all measured at $T=50\,\mathrm{mK}$. The insets demonstrate the corresponding $I$-$V_{\mathrm{sd}}$ curves at the specified $V_{\text{g}}$ values. (d-f) The same measurements for the $38\,\mathrm{{\mu}m}$-long edge.}
	\label{hgte_additional_8nm_par}
\end{figure*}

\newpage
\begin{figure*}[h]
	\begin{center}
		\includegraphics[scale=1]{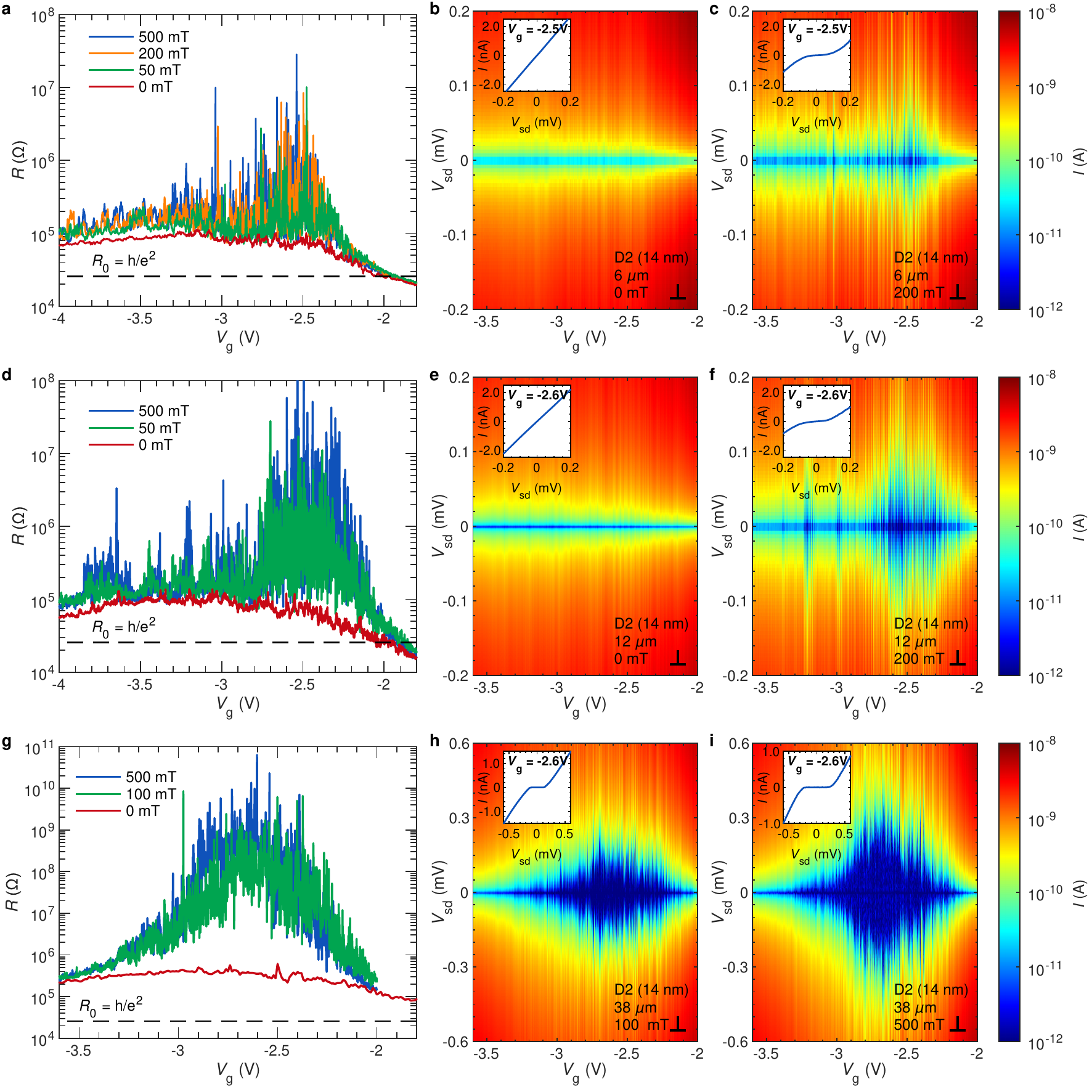}
	\end{center}
	\caption{\textbf{The effect of perpendicular magnetic field in the $\mathbf{14\,\mathbf{nm}}$ QW.} (a) The two-terminal $R(V_{\text{g}})$-dependence of a $6\,\mathrm{{\mu}m}$-long edge in~D2 at various $B_{\perp}$ and (b,c) the gap-opening in the same edge at $B=200\,\mathrm{mT}$, all measured at $T=50\,\mathrm{mK}$. The insets demonstrate the corresponding $I$-$V_{\mathrm{sd}}$ curves at the specified $V_{\text{g}}$ values. (d-f) The same measurements for the $12\,\mathrm{{\mu}m}$-long edge. (g-i) Similar measurements for the $38\,\mathrm{{\mu}m}$-long edge with panels (h,i)  demonstrating the change in the gap with increasing magnetic field.}
	\label{hgte_additional_14nm}
\end{figure*}

\newpage
\begin{figure*}[h]
	\begin{center}
		\includegraphics[scale=1]{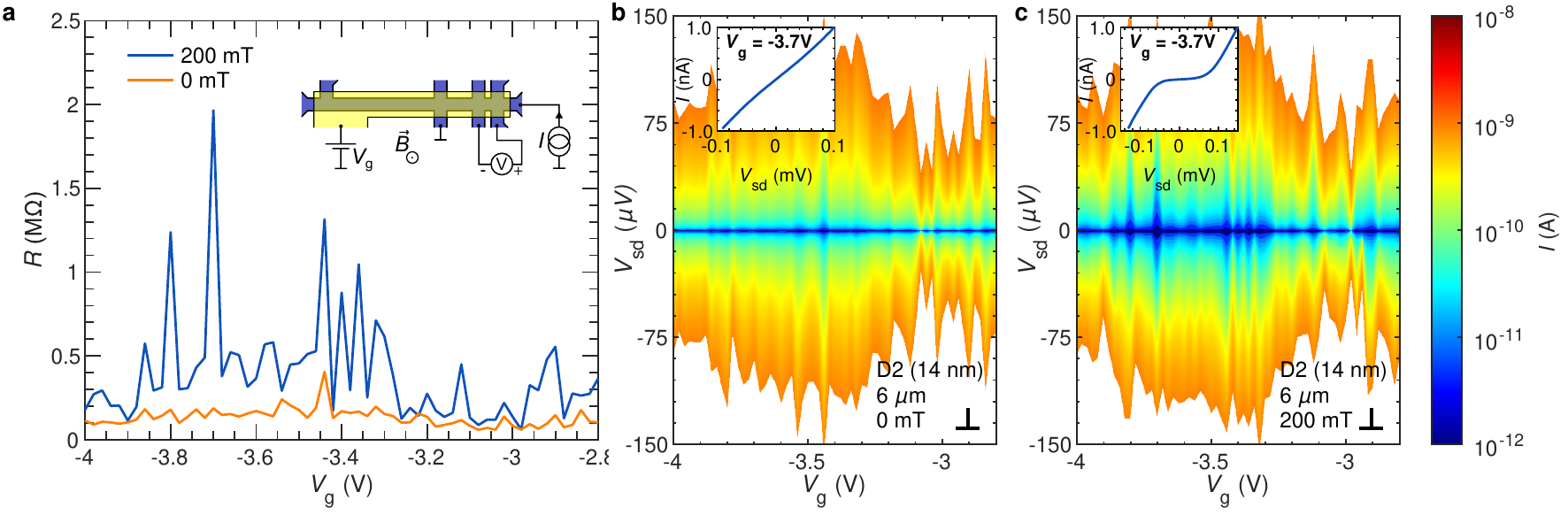}
	\end{center}
	\caption{\textbf{The effect of perpendicular magnetic field in the $\mathbf{14\,\mathbf{nm}}$ QW.} (a) The four-terminal $R(V_{\text{g}})$-dependence of a $6\,\mathrm{{\mu}m}$-long edge in~D2 at $B_{\perp}=0\text{ and }200\,\mathrm{mT}$ and (b,c) the gap-opening in the same edge at $B=200\,\mathrm{mT}$, all measured at $T=50\,\mathrm{mK}$. The insets demonstrate the corresponding $I$-$V_{\mathrm{sd}}$ curves at the specified $V_{\text{g}}$ values.}
	\label{hgte_additional_14nm}
\end{figure*}
\end{document}